\documentclass[12pt]{article}
\usepackage{a4wide,epsfig,amssymb}
\newcommand{\gnorst}{\ensuremath{\alpha_S}}
\newcommand{\strong}{\ensuremath{\bar\gnorst}}
\newcommand{\colour}{\left({C_A\over C_F}\right)}
\newcommand{\strLsq}{\left(\strong L^2\right)}
\newcommand{\CFstrLsq}{\left(C_F\strong L^2\right)}
\newcommand{\CAstrLsq}{\left(C_A\strong L^2\right)}
\newcommand{\open}{&\sim&\left[\vphantom{\colour^2}}
\newcommand{\close}{\right]\CFstrLsq}
\newcommand{\openstrLsq}{&\sim&\vphantom{\strLsq^2}}
\newcommand{\openfraction}{&\sim&\vphantom{\colour^2}}
\begin{document}

\thispagestyle{empty}
\begin{center}
\vspace{2cm}

\begin{flushright}
Cavendish-HEP-96/21

hep-ph/9610552

31 October 1996
\end{flushright}
\vspace{2cm}

{\Large
        {\bf
        Jet Fractions in \boldmath{$e^+e^-$} Annihilation
        }
\vspace{2cm}

Garth Leder
\footnotemark
}
\vspace{5mm}

{\it
Cavendish Laboratory,
Cambridge University,

Madingley Road,
Cambridge CB3 0HE.
ENGLAND
}

\vspace{5mm}

email: leder@hep.phy.cam.ac.uk

\vspace{2cm}

{\bf\large
Abstract
}
\vspace{5mm}

\parbox{10cm}{
  
  The jet fractions expected in $e^+e^-$ annihilation are calculated
  analytically in the leading log approximation up to 6 jet order for
  both the JADE and Durham algorithms.

}

\footnotetext{
This research was supported in part by the UK Particle Physics and
Astronomy Research Council.
}

\end{center}

\clearpage

\tableofcontents
\clearpage

\section{Introduction}

\subsection{QCD and Jet Fractions}

QCD is now well accepted as the theory of strong interactions.  It has
a running coupling constant $\strong={\gnorst(Q^2) / 2\pi}$, which is
only low enough for perturbation theory to be valid at high energies:
$Q^2$ is the relevant momentum scale.  In the LEP and SLC $e^+e^-$
colliders, the centre-of-mass energy $Q\gtrsim 90$ GeV is high enough
for perturbation theory to be valid in the early stages of the strong
interaction~\cite{jets/LEP_energies}: it can therefore be used to
calculate the production of quarks and gluons (partons).  These
energies also allow the determination of the QCD colour factors $C_F$
and $C_A$~\cite{QCD/colour_factors/ALEPH, QCD/colour_factors/OPAL}.
The partons then combine at lower momentum scales where perturbation
theory is not valid into a large number of hadrons in the process
known as hadronisation.  It is these hadrons that are actually
measured in the detectors.

\label{LPHD}

The distribution of experimentally measured hadrons closely follows
the theoretically calculated partons in its flow of momentum and
quantum numbers~\cite{Perturbative_QCD, LPHD}.  The uncertainty
principle relates the distance over which an interaction takes place
to the momentum transfer involved.  Interactions with a momentum scale
greater than about $1$ GeV are in the perturbative regime.  In the
non-perturbative regime there are large distances between partons, and
the momentum transfer at this stage is therefore small.  This
phenomenon, known as Local Parton-Hadron Duality, allows hadronisation
to be modelled by Monte Carlo event generators such as
JETSET~\cite{JETSET/paper1, JETSET/paper2} and
HERWIG~\cite{HERWIG,HERWIG/simulation1, HERWIG/simulation2}.  The
former uses a string model in which the gluon field energy is
converted into hadrons, whilst the latter uses a low-mass cluster
model.  These two event generators yield similar results, for example
in the determination of $\gnorst$~\cite{OPAL/alpha_s/determination,
ALEPH/alpha_s/determination, L3/alpha_s/determination,
DELPHI/alpha_s/determination}, giving some confidence in their
validity.

\label{algorithm_definitions}

One of the common observables used in the study of hadronic final
states is the number of jets as defined by a given jet algorithm.  The
$n$th jet fraction $R_n$ is then defined as
\begin{equation}
R_n={\sigma_n \over \sigma_{total}}
\end{equation}
where $\sigma_n$ is the cross section for $n$ jets and
$\sigma_{total}$ is the total cross section.  Jet fractions may be
used to calculate $\gnorst$ by comparing parton-level theoretical
calculations with measured hadron-level rates, provided the same jet
algorithm is used in the two stages of the analysis.

Jet algorithms use a measure to determine the closeness of pairs of
tracks.  The JADE algorithm~\cite{JADE/algorithm/paper1,
JADE/algorithm/paper2} uses the invariant mass
\begin{equation}
y^J_{ij}={(p_i+p_j)^2 \over Q^2}={2p_ip_j \over Q^2}
\end{equation}
where $p_i$ and $p_j$ are the four-momenta of the two particles
(assumed massless) and $Q^2$ is the centre-of-mass energy squared of
the complete event as defined above.  Let the partons have energies
$E_i$, $E_j$ and relative angle $\theta$ in the centre-of-mass frame.
The measure is then
\begin{equation}
y^J_{ij}={2E_iE_j(1-\cos\theta) \over Q^2}.
\label{jade_measure}
\end{equation}
At each stage, the pair of tracks with the smallest measure are
combined if this measure is smaller than a fixed cut-off, ie if
$y_{ij}<y_{cut}$.  The algorithm stops once all pairs have a measure
greater than $y_{cut}$, at which point the remaining tracks are termed
jets.  In experimental analyses the tracks are hadrons in the
detector; in theoretical calculations they are partons.  The validity
of comparing these two situations is at least suggested by Local
Parton-Hadron Duality, and might provide information about
non-perturbative effects.

The Durham algorithm~\cite{jets/moment_generating/Durham} is identical
to the JADE algorithm except that it uses as its measure
\begin{equation}
y^D_{ij}={2\,{\rm min}\{E_i^2,E_j^2\}(1-\cos\theta) \over Q^2}.
\label{durham_measure}
\end{equation}
A further variant on this theme is the Geneva
algorithm~\cite{jets/algorithms/NLO}, again identical except for its
measure:
\begin{equation}
y^G_{ij}={8 \over 9}{E_iE_j(1-\cos\theta) \over (E_i+E_j)^2}.
\end{equation}
This algorithm has not been widely used in experimental analyses.  For
each measure there are a number of different possible procedures for
combining pairs of tracks, the simplest of which is to add their
four-momenta.  The calculations presented in this paper are of an
accuracy not to be affected by the procedure used.

\label{parton_masses}

The mass of a current quark ($\sim$ few
MeV)~\cite{quark_masses/review} is much less than the characteristic
QCD scale ($\Lambda_{QCD}\simeq~200$ MeV).  For the above algorithms
the effect of the masses is swamped by non-perturbative effects
\cite{quark_masses/jets}, and the quarks may be considered massless.

\label{angular_ordering}

At small values of $y_{cut}$ the jet fractions are dominated by soft
gluons, which have a useful property known as angular ordering.
Consider the emission of such a soft gluon from either of a pair of
partons separated by an angle $\theta$.  The emitting partons may be
treated as classical currents: the emission from each of these
currents interferes, but is coherent due to the colour structure.  The
resulting cross section \cite{jets/monte_carlo}, written in terms of
the energies and angles of the partons, may be split into two parts,
each containing a collinear singularity.  When averaged over the
azimuthal direction the contribution from each of these parts is
non-zero only within a cone of angle $\theta$: the emission may be
separated into two parts, each of which describes emission from a
single parton, but restricted to the cone described above.  Emission
at larger angles adds up to be equivalent to emission from the parent
parton~\cite{coherence/review}.  This argument holds to
next-to-leading log order, and was discovered independently by
Mueller~\cite{coherence/discovery1} and Ermolaev and
Fadin~\cite{coherence/discovery2}.  Coherence effects of this type
have been observed at a variety of
energies~\cite{coherence/observation1, coherence/observation2,
coherence/observation/OPAL}.

\subsection{The Leading Logarithmic Approximation}

\begin{figure}
\begin{center}
\epsfig{file=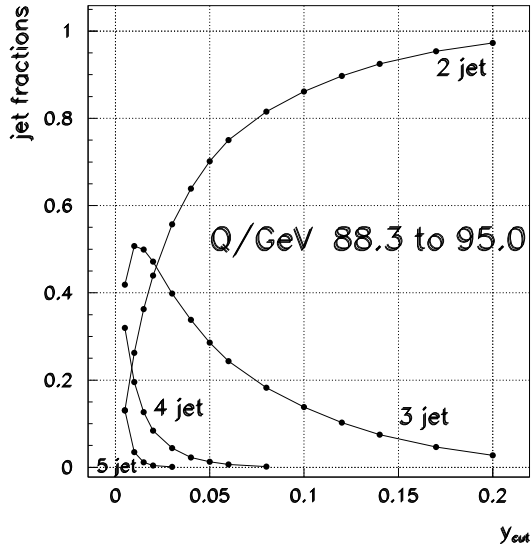, width=0.5\textwidth}
\caption[Jet rates at OPAL]{
Jet rates as measured by the OPAL collaboration.  Data taken
from~\cite{OPAL/recombination_schemes} }
\label{jet_rates}
\end{center}
\end{figure}

Hadronisation corrections set in at $y_{cut} \sim m^2_{Hadrons}/Q^2
\sim (1$ GeV$)^2/Q^2$: at LEP energies this gives a lower limit on
$y_{cut}$ of $\sim 0.0001$.  At values of $y_{cut}<0.1$, such that
\begin{equation}
L=\ln\left({1 \over y_{cut}}\right)>2.5,
\end{equation}
the divergent regions of phase space due to soft, collinear gluons
dominate the integrand.  Perturbation theory then gives a series in
two variables, in increasing powers of $\strong$ and decreasing powers
of $L$.  For jet fractions the leading term in $L$ has two powers of
$L$ for every power of $\strong$, one from each of the soft and
collinear divergences.  At each order of $\strong$ the series in $L$
can be truncated to give a valid approximation, provided that $\strong
L^2<1$.  I shall work to leading log order, in which only the highest
power of $L$ is kept.  At LEP energies $\strong\simeq 0.117 \pm 0.005$
\cite{Glasgow_talk}, giving a lower limit on $y_{cut}$ of $\sim 0.05$.  At
higher energies this limit will be lowered, widening the valid range
of $y_{cut}$.

At lower values of $y_{cut}$ at LEP energies it is necessary to resum
the series to all orders in $\strong$.  This is possible if the
coefficients follow, or are at least close to, the pattern of a known
function; for example the leading log coefficients for the two jet
fraction in the Durham algorithm follow the pattern of the
coefficients of the exponential
function~\cite{jets/moment_generating/500GeV}.  This phenomenon is
known as exponentiation~\cite{exponentiation}, and depends on the
factorisation of the phase space of the relevant observable: the QCD
matrix elements are guaranteed to factorise for soft, collinear
gluons~\cite{matrix_element_factorisation}.  These resummed
predictions compare well with data~\cite{jets/experimental_results},
and may be used to extract the value of $\strong$, the main error
being theoretical.

The same properties do not hold for the JADE algorithm, which has more
complicated phase space boundaries due to extra gluon-gluon
correlations: the coefficients follow no discernable pattern for the
most commonly used recombination
prescriptions~\cite{jets/moment_generating_introduction}.

Calculations for jet fractions exist at leading order in $\strong$ and
leading log order up to six jet order for the JADE
algorithm~\cite{jet_fractions/JADE}, and up to four jet order for the
Durham algorithm~\cite{jets/moment_generating/Durham,
jet_fractions/Durham}.  The JADE calculations only include the abelian
terms, that is only the diagrams with no triple gluon vertices.  The
existing results are here extended to include the non-abelian terms.
The existence of the triple gluon vertex has been confirmed at
LEP~\cite{QCD/colour_factors/ALEPH}.  Some of the coefficients have
also been investigated numerically~\cite{jets/large_logs}.

Jet clustering has also been used as a first stage in experimental
analyses to reduce the theoretical errors in other
observables~\cite{ALEPH/clustering}.  The calculation of jet fractions
can also be used as a test of the consistency of QCD between different
experimental situations, for example between $e^+e^-$ annihilation and
deep inelastic scattering.

\subsection{Soft and Collinear Divergences}

In the perturbative calculation of parton branching there are two
types of divergence that occur even aside from renormalisation.  In
the case of branching from an outgoing parton (time-like branching),
and after integrating over the azimuthal direction, the cross section
can be expressed as~\cite{QCD_and_Collider_Physics}
\begin{equation}
{d\sigma_{n+1} \over d\sigma_n}=\strong V(z)dz
{\sin\theta d\theta \over 1-\cos\theta}
\label{evolution}
\end{equation}
where $\theta$ is the angle between the partons.  For $q\rightarrow
qg$ branching $z$ is the energy fraction of the emitted gluon, whilst
for $g\rightarrow gg$ branching it is the energy fraction of either
gluon.  The formulation in terms of parton energies and angles is used
in order to take into account both types of divergence.  $V(z)$ is the
splitting function which gives the probability of parton branching
after averaging and summing over initial and final spin states
respectively.  In the soft (small $z$) limit the two splitting
functions relevant to this analysis~\cite{jets/monte_carlo} are
\begin{equation}
V_F^G(z) [q\rightarrow qg] = C_F\,{1+(1-z)^2 \over z} \simeq {2C_F \over z} +
{\cal O}(1)
\end{equation}
and
\begin{equation}
V_G^G(z) [g\rightarrow gg] = 2C_A\left[z\,(1-z)+ {1-z \over z} + 
{z \over 1-z }\right] \simeq {2C_A \over z} + {\cal O}(1).
\end{equation}
These splitting functions diverge at $z=0$.  The evolution equation
(\ref{evolution}) is also divergent in the collinear (small $\theta$)
limit:
\begin{equation}
{d\sigma_{n+1}\over d\sigma_n} \simeq 2\strong V(z)dz
{d\theta\over\theta} \simeq 2\strong C {dz \over z}
{d\theta^2 \over \theta^2},
\label{phase_space}
\end{equation}
where $C=C_F$ or $C_A$ according to the type of branching.  This
divergence is a consequence of the partons' masslessness, but the
expression would be logarithmically enhanced even with small parton
masses.  The splitting function for gluon to quark-antiquark branching
is
\begin{equation}
V_G^F(z) [g\rightarrow qq] = {1 \over 2}N_f\left[z^2+(1-z)^2\right],
\end{equation}
where $N_f$ is the number of active quark flavours at the relevant
energy.  This function contains no soft divergence, and therefore this
type of branching does not contribute at leading log order.  Similarly
the four gluon vertex is suppressed by a factor of $\strong$.

For an observable to be physical it must not be affected by any of
the divergences in the theory.  It must therefore take the same value
if any parton is replaced by an arbitrary number of partons such that
the total momentum and colour flow of the group remains the same.
Such an observable is called collinear-safe.  For the same reason it
must also remain unchanged if a parton emits an arbitrary number of
gluons of zero energy, in which case it is called infrared-safe.  This
suggests that a parton is significant for its momentum and colour
flow, rather than as a directly measurable particle.

\section{Analytic Calculation of Jet Fractions}
\label{analytic}

For the leading log order result all the gluons are soft and
collinear.  The soft limit enables the energy fraction of each parton
to be approximated as $\epsilon_i\simeq E_i/Q$.  Note that the
original quark and antiquark both have $\epsilon=1$ to lowest order in
$\epsilon$.  For two partons branching off the same initial parton the
collinear limit ($\theta$ small) allows the approximation
$2(1-\cos\theta) \simeq \theta^2 + {\cal O}(\theta^4)$ to be made.
The measures (\ref{jade_measure}) and (\ref{durham_measure}) then
simplify to
\begin{eqnarray}
y^J_{ij}&=&\epsilon_i\epsilon_j\theta^2, \\
y^D_{ij}&=&{\rm min}\{\epsilon_i^2,\epsilon_j^2\}\theta^2.
\end{eqnarray}
For two gluons originating one from the quark and one from the
antiquark, the collinear limit is $\theta \simeq \pi$ and the measures
become
\begin{eqnarray}
y^J_{ij}&=&\epsilon_i\epsilon_j, \\
y^D_{ij}&=&{\rm min}\{\epsilon_i^2,\epsilon_j^2\}.
\end{eqnarray}
These approximations are sufficient for the leading log order
calculation.  In the small $\theta$ limit the perpendicular momentum
of the less energetic particle with respect to the more energetic
particle is $k_\perp=Q\epsilon\sin\theta\simeq Q\epsilon\,\theta$
where $Q\epsilon$ is the energy of the less energetic particle as
defined above.  The Durham measure is thus seen to be equal to the
perpendicular momentum squared divided by $Q^2$.

The phase space for $n$ jets (\ref{phase_space}) is
\begin{equation}
{d\epsilon_1 \over \epsilon_1}...{d\epsilon_n \over
\epsilon_n}{d\theta^2_1 \over \theta^2_1}...{d\theta^2_n \over \theta^2_n}.
\end{equation}
The substitutions
\begin{equation}
\ln\left(\vphantom{1\over y_{cut}}1\over\epsilon_i\right)=x_iL,
\quad\ln\left(\vphantom{1\over y_{cut}}1\over\theta_i^2\right)=y_iL,
\quad\ln\left(\vphantom{1\over y_{cut}}1\over y_{cut}\right)=L,
\label{substitutions}
\end{equation}
transform this to
\begin{equation}
dx_1...dx_ndy_1...dy_n.
\end{equation}
The formal integration range for the original variables
$\epsilon_i,\theta^2_i$ is taken as $0 \rightarrow 1$, remembering
that only small values of $\epsilon$ and $\theta$ contribute at
leading log order.  This translates into $x_i,y_i$ space as $0
\rightarrow \infty$.  In the actual calculation it is simplest to put
in these limits as step functions, ie
$\Theta(x_1)...\Theta(x_n)\Theta(y_1)...\Theta(y_n)$, where
\begin{equation}
\Theta(x)=\cases{1 & $x \ge 0$, \cr 0 & $x < 0$.}
\end{equation}

I will work for the moment with the JADE algorithm.  Consider two
gluons radiated successively by a quark with energy fractions
$\epsilon_i$ and $\epsilon_j$, and at angles $\theta_i$ and $\theta_j$
respectively to the quark.  In order to generate the maximum number of
logarithms $L$ the angles must be strongly ordered, ie
$\theta_j\ll\theta_i$~\cite{jets/monte_carlo}.  The smaller of the
angles ($\theta_j$) may therefore be ignored, and the angle between
the two partons taken as $\theta_i$.  The condition that these two
gluons are resolved as separate jets is $y_{ij}>y_{cut}$, which
appears in the integrand as
\begin{equation}
\Theta(\epsilon_i\epsilon_j\theta_{i}^2-y_{cut});
\label{step}
\end{equation}
the substitutions (\ref{substitutions}) transform this to
\begin{equation}
\Theta(1-x_i-x_j-y_i).
\end{equation}
This divides the phase space into two halves separated by a plane.
Taken together with the appropriate conditions for other pairs of
partons the overall integration region is a `polygon' in a space with
two dimensions for every emitted gluon.

For the specific diagram
\begin{center}
\epsfig{file=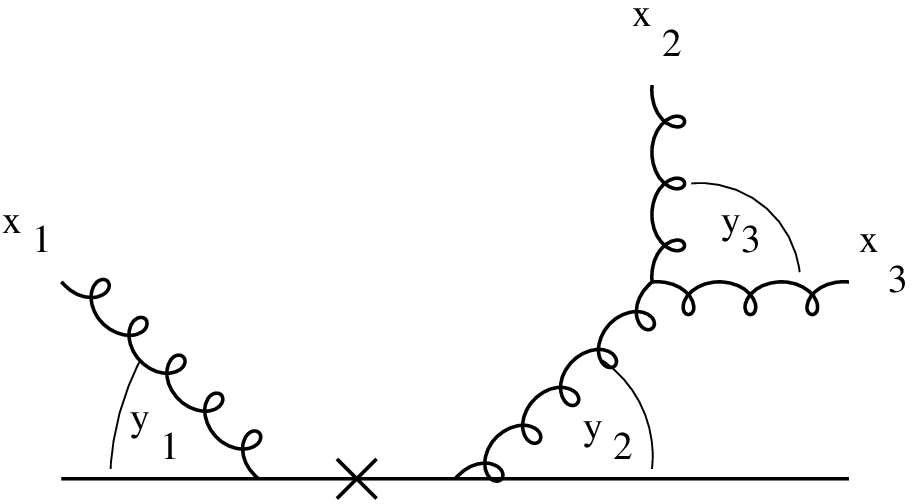}
\end{center}
the conditions that no two partons will be combined by the JADE
algorithm are
\begin{center}
\begin{tabular}{|l|l|} \hline
angular ordering & $\Theta(\theta_2-\theta_3)$ \\ \hline
quark-gluon & $\Theta(\epsilon_1\theta_1^2-y_{cut})$ \\ \hline
gluon-gluon & $\Theta(\epsilon_1\epsilon_2-y_{cut})$ \\ \cline{2-2}
& $\Theta(\epsilon_1\epsilon_3-y_{cut})$ \\ \cline{2-2}
& $\Theta(\epsilon_2\epsilon_3\theta_3^2-y_{cut})$ \\ \hline
\end{tabular}
\end{center}
This gives as the integration
\begin{eqnarray}
R&=&\int_{-\infty}^{\infty}\Theta(y_3-y_2)\Theta(1-x_1-y_1)
\Theta(1-x_1-x_2)\Theta(1-x_1-x_3)\Theta(1-x_2-x_3-y_2) \cr && \qquad
\Theta(x_1)\Theta(x_2)\Theta(x_3)\Theta(y_1)\Theta(y_2)\Theta(y_3)
dx_1dx_2dx_3dy_1dy_2dy_3.
\end{eqnarray}
This calculation may be done using the C program outlined in appendix
(\ref{program_details}): the answer is $53/360$, which yields a
contribution of
\begin{equation}
{1 \over 2}2^3C_F^2C_A{53 \over 360}\strong^3L^6={424 \over
135}\strLsq^3
\end{equation}
to the final answer.  The factor of $1/2$ is statistical, and results
from gluons 2 and 3 being identical.  In this context a pair of gluons
are identical if the swapping of their coordinates leaves the
integrand invariant: such gluons must either contribute a statistical
factor of $1/2$ or have their energies ordered.  To reach physically
meaningful results it is necessary to add up the results from each of
the diagrams contributing at the relevant order.  At leading log order
for $n$ jets this consists of all the tree level diagrams involving
the original quark-antiquark pair and $n-2$ emitted gluons; the
creation of further quark-antiquark pairs is suppressed by one factor
of $L$ as there is no associated soft divergence.

It is of course possible to integrate over the variables in any order.
The accuracy of the program had been tested for a number of
integrands: it gives the same answer for several different integration
orders.  The results presented in this paper have been found using at
least two different integration orders, always with consistent
results.  All the results have also been checked numerically, using
the VEGAS Monte Carlo routine in Fortran 77.  Here the integration was
performed directly in $\epsilon,\theta$ space.  Again the results are
consistent within the errors of the Monte Carlo.

\section{Moment Generating Function Method}
\label{moment}

The moment generating function for a probability distribution $p(n)$
is defined as
\begin{equation}
\phi(u)=\sum_{n=0}^\infty u^n p(n).
\end{equation}
The probabilities themselves are recovered using the formula
\begin{equation}
p(n)={1 \over n!}\left[{d^n \phantom{u} \over
du^n}\left(\phi(u)\right)\right]_{u=0}.
\label{recover}
\end{equation}
For the case of a $q \bar q$ pair produced in $e^+e^-$ annihilation,
the moment generating function for the Durham algorithm may be
calculated to next-to-leading order using the coherent branching
formalism \cite{resummation/event_shapes}: it is
$\phi_q^2(Q,u)=u^2e^{2F_0(Q,u)}$,
where~\cite{jets/moment_generating/Durham}
\begin{eqnarray}
F_i(Q,u)&=&{d^i \phantom{u} \over du^i}
\int_{Q_0}^Q\Gamma_q(Q,q)\left[\phi_g(q,u)-1\right]dq \cr
&=&\int_{Q_0}^Q\Gamma_q(Q,q)\left[{d^i \phantom{u} \over du^i}
\phi_g(q,u)\right]dq,
\quad i\ge 1.
\end{eqnarray}
Here
\begin{equation}
\Gamma_q(Q,q)={4C_F\strong \over q}\left[\ln\left({Q \over 
q}\right)-{3 \over 4}\right]
\end{equation}
is the $q\rightarrow qg$ branching probability, and $\phi_g(q,u)$ is
the moment generating function for a single gluon.  At leading log
order $\strong$ may be taken as a constant, evaluated at the momentum
scale of the interaction.  Defining
\begin{equation}
G_i(Q,u)={d^i \phantom{u} \over du^i}\left[\phi_q^2(Q,u)\right],
\end{equation}
it may easily be calculated that
\begin{eqnarray}
G_0&=&\left[u^2\right]e^{2F_0} \cr
G_1&=&\left[2u+2u^2F_1\right]e^{2F_0} \cr
G_2&=&\left[2+8uF_1 + u^2\left(2F_2+4F_1^2\right)\right]e^{2F_0} \cr
G_3&=&\left[12F_1 + u\left(12F_2+24F_1^2\right) +
u^2\left(2F_3+12F_1F_2+8F_1^3\right)\right]e^{2F_0} \cr
G_4&=&\left[24F_2+48F_1^2 + u\left(16F_3+96F_1F_2+64F_1^3\right)\right. \cr
&& \left. {} +
u^2\left(2F_4+16F_1F_3+12F_2^2+48F_1^2F_2+16F_1^4\right) \right]e^{2F_0}
\cr
G_5&=&\left[40F_3+240F_1F_2+160F_1^3 +
u\left(20F_4+160F_1F_3+120F_2^2\right.\right. \cr
&& \left.\left. {}+480F_1^2F_2+160F_1^4\right) + {\cal
O}\left(u^2\right)\right]e^{2F_0}
\cr
G_6&=&\left[60F_4+440F_1F_3+360F_2^2+1200F_1^2F_2+320F_1^4 + {\cal
O}\left(\vphantom{u^2}u\right)\right]e^{2F_0},
\end{eqnarray}
where the $G_i$ and $F_i$ are evaluated at $(Q,u)$.  The quark Sudakov
form factor is calculated as
\begin{equation}
\Delta_q(Q)=e^{F_0(Q,0)} = e^{-\int_{Q_0}^Q\Gamma_q(Q,q)dq}
\end{equation}
using $\phi_g(q,0)=0$ as calculated in the next section (\ref{A_0=0}).
The jet rates themselves are found using equation (\ref{recover}) to
be
\begin{eqnarray}
R_2(Q)&=&\vphantom{\left(F_1^2\right)}\Delta_q^2(Q) \cr
R_3(Q)&=&\left(\vphantom{F_1^2}2F_1\right)\Delta_q^2(Q) \cr
R_4(Q)&=&\left(F_2+2F_1^2\right)\Delta_q^2(Q) \cr
R_5(Q)&=&\left(\mbox{$1 \over 3$}F_3+2F_1F_2+\mbox{$4 \over
3$}F_1^3\right)\Delta_q^2(Q) \cr
R_6(Q)&=&\left(\mbox{$1 \over 12$}F_4+\mbox{$11 \over 18$}F_1F_3+\mbox{$1
\over 2$}F_2^2+\mbox{$5
\over 3$}F_1^2F_2+\mbox{$4 \over 9$}F_1^4\right)\Delta_q^2(Q),
\end{eqnarray}
where the $F_i$ are evaluated at $(Q,0)$.

The gluon moment generating function
is~\cite{jets/moment_generating/Durham}
\begin{equation}
\phi_g(Q,u)=\left[u+u^2D_0(Q,u)\right]e^{C_0(Q,u)},
\label{gmgf1}
\end{equation}
where
\begin{equation}
D_i(Q,u)={d^i \phantom{u} \over du^i}
\int_{Q_0}^Q \Gamma_f(q)e^{E_0(q,u)}dq=\int_{Q_0}^Q\Gamma_f(q)
\left[{d^i \phantom{u} \over du^i}e^{E_0(q,u)}\right]dq,
\end{equation}
\begin{eqnarray}
E_i(Q,u)&=&{d^i \phantom{u} \over du^i} \int_{Q_0}^Q
\left\{\left[2\Gamma_q(Q,q)-\Gamma_g(Q,q)\right]
\left[\phi_g(q,u)-1\right]+\Gamma_f(q) \right\} dq \cr
&=& \int_{Q_0}^Q \left[2\Gamma_q(Q,q)-\Gamma_g(Q,q)\right] \left[{d^i
\phantom{u} \over du^i} \phi_g(q,u)\right]dq, \quad i\ge 1,
\end{eqnarray}
and
\begin{eqnarray}
C_i(Q,u)&=&{d^i \phantom{u} \over du^i} \int_{Q_0}^Q \left\{\Gamma_g(Q,q)
\left[\phi_g(q,u)-1\right]-\Gamma_f(q) \right\} dq \cr
&=&\int_{Q_0}^Q\Gamma_g(Q,q)\left[{d^i \phantom{u} \over du^i}
\phi_g(q,u)\right]dq, \quad i\ge 1.
\label{gmgf2}
\end{eqnarray}
Here
\begin{equation}
\Gamma_g(Q,q)= {4C_A\strong \over q} \left[\ln\left({Q \over q}\right)
- {11 \over 12}\right]
\label{Gamma_g}
\end{equation}
and
\begin{equation}
\Gamma_f(q)={2N_f\strong \over 3q}
\end{equation}
are $g\rightarrow gg$ and $g\rightarrow q\bar q$ branching
probabilities respectively.  Defining
\begin{equation}
A_i(Q,u)={d^i \phantom{u} \over du^i}
\left[\vphantom{\phi_q^2}\phi_g(Q,u)\right],
\end{equation}
it may easily be calculated that
\begin{eqnarray}
A_0(Q,u)&=&\left[u+u^2D_0\right]e^{C_0} \cr
A_1(Q,u)&=&\left[1 + u\left(C_1+2D_0\right) +
u^2\left(D_1+D_0C_1\right)\right]e^{C_0} \cr
A_2(Q,u)&=&\left[2C_1+2D_0 +
u\left(C_2+C_1^2+4D_1+4D_0C_1\right)\right. \cr
&& \left. {}+ u^2\left(D_2+2D_1C_1+D_0C_2+D_0C_1^2\right)\right]e^{C_0}
\cr
A_3(Q,u)&=&\left[3C_2+3C_1^2+6D_1+6D_0C_1 +
u\left(C_3+3C_1C_2+C_1^3+6D_2\right.\right. \cr
&& \left.\left. {}+12D_1C_1+6D_0C_2+6D_0C_1^2\right) + {\cal
O}\left(u^2\right) \right]e^{C_0} \cr
A_4(Q,u)&=&\left[4C_3+12C_1C_2+4C_1^3+12D_2+24D_1C_1+12D_0C_2+12D_0C_1^2
\right. \cr
&& \left. \vphantom{C_1^2} {}+ {\cal O}\left(\vphantom{u^2}u\right)
\right]e^{C_0},
\end{eqnarray}
where the $A_i$, $C_i$ and $D_i$ are evaluated at $(Q,u)$.  The gluon
Sudakov form factor is calculated as
\begin{equation}
\Delta_g(Q)=e^{C_0(Q,0)}=e^{-\int_{Q_0}^Q\Gamma_g(Q,q)+\Gamma_f(q)dq}
\end{equation}
again using $\phi_g(q,0)=0$.  The $A_i$ are evaluated at $(Q,0)$ to be
\begin{eqnarray}
A_0(Q,0) &=& \vphantom{\left(C_1^2\right)} 0 \cr
A_1(Q,0) &=& \vphantom{\left(C_1^2\right)} \Delta_g(Q) \cr
A_2(Q,0) &=& 2\left( \vphantom{C_1^2} C_1+D_0\right) \Delta_g(Q) \cr
A_3(Q,0) &=& 3\left(C_2+C_1^2+2D_1+2D_0C_1\right) \Delta_g(Q) \cr
A_4(Q,0) &=& 4\left(C_3+3C_1C_2+C_1^3+3D_2+6D_1C_1+3D_0C_2+3D_0C_1^2
\right)\Delta_g(Q),
\label{A_0=0}
\end{eqnarray}
where the $C_i$ and $D_i$ are also evaluated at $(Q,0)$.  Defining a
further form factor as
\begin{equation}
\Delta_f(Q) = e^{E_0(Q,0)} = e^{-\int_{Q_0}^Q 2\Gamma_q(Q,q)-\Gamma_g(Q,q)
-\Gamma_f(q)dq} = {\Delta_q^2(Q) \over \Delta_g(Q)},
\end{equation}
and defining
\begin{equation}
B_i(Q,u)={d^i \phantom{u} \over du^i}\left[e^{E_0(Q,u)}\right],
\end{equation}
it may similarly be calculated that
\begin{eqnarray}
B_0(Q,0) &=& \vphantom{\left(D_0^2\right)} \Delta_f(Q) \cr
B_1(Q,0) &=& \left( \vphantom{D_0^2} E_1 \right) \Delta_f(Q) \cr
B_2(Q,0) &=& \left( E_2+E_1^2 \right) \Delta_f(Q),
\end{eqnarray}
where the $E_i$ are also evaluated at $(Q,0)$.

\begin{figure}
\begin{center}
\begin{tabular}{|l|l|} \hline
Quantity & Depends on \\ \hline
$A_n$ & $C_{0..n-1}, D_{0..n-2}$ \\ \hline
$E_n$ & $A_n$ \\ \hline
$B_n$ & $E_{0..n-1}$ \\ \hline
$C_n$ & $A_n$ \\ \hline
$D_n$ & $B_n$ \\ \hline
\end{tabular}
\caption[]{Quantity dependencies in the moment generating function method}
\label{dependencies}
\end{center}
\end{figure}

Figure~(\ref{dependencies}) shows the dependencies of each of the
quantities $A_i$ to $E_i$.  The quantities may be calculated
cyclically in the order in which they are given.  The key point is
that $A_n$ only depends (explicitly or implicitly) on $A_{0..n-1}$ at
$u=0$, and so the recursive nature of equations (\ref{gmgf1}) to
(\ref{gmgf2}) is avoided.  It is interesting to note how the colour
factors are introduced by the different quantities: for example each
factor of $C_i(Q,0)$ introduces a colour factor of $C_A$ through its
dependency on $\Gamma_g(Q,q)$ (equation (\ref{Gamma_g})).

\section{Results}

$C_F$ and $C_A$ are the QCD colour factors, $\strong=\gnorst/2\pi$ and
$L=\ln(1/y_{cut})$.  $R^J_n$ and $R^D_n$ are the leading log order $n$
jet rates for the JADE and Durham algorithms respectively.  The JADE
results were found by direct integration as described in
section~(\ref{analytic}), and the Durham results using both that and
the moment generating function method described in
section~(\ref{moment}): the two methods agree at all calculated
orders.  In addition, as mentioned earlier, the results for both
algorithms have been checked numerically using the Monte Carlo
Integrator VEGAS, again with agreement within the errors of the Monte
Carlo.
\begin{eqnarray}
R^J_2\open 1\right]\cr
R^J_3\open 2\close\cr
R^J_4\open {3\over 2}+{1\over 6}\colour\close ^2\cr
R^J_5\open {31\over 45}+{5\over 18}\colour+{1\over 36}\colour^2\close
^3\cr
R^J_6\open {571\over 2520}+{4091\over 20160}\colour+{643\over
11520}\colour^2+{107\over 20160}\colour^3\close ^4\cr
R^D_2\open 1\right]\cr
R^D_3\open 1\close\cr
R^D_4\open {1\over 2}+{1\over 12}\colour\close ^2\cr
R^D_5\open {1\over 6}+{1\over 12}\colour+{1\over 90}\colour^2\close
^3\cr
R^D_6\open {1\over 24}+{1\over 24}\colour+{7\over
480}\colour^2+{17\over 10080}\colour^3\close ^4.
\label{main_results}
\end{eqnarray}
An additional result is the abelian contribution to the 7 jet fraction
for the two algorithms:
\begin{eqnarray}
R^J_7\open{1093\over 18900}\close ^5,\cr
R^D_7\open{1\over 120}\close ^5.
\end{eqnarray}
Whilst these abelian contributions cannot be used on their own, they
provide further evidence that the JADE coefficients do not follow an
obvious pattern, while the Durham abelian coefficients exponentiate.

One qualitative aspect of the Durham algorithm is that soft, collinear
gluons are combined only such that at least one of the pair has no
other partons closer to it in angle than its partner.  This suggests
that the phase space boundaries given by comparing other pairs of
gluons should have no effect on the result: this was confirmed for all
the results quoted above.

These results are in complete agreement with the results of Brown and
Stirling as presented in \cite{jet_fractions/JADE} and
\cite{jet_fractions/Durham} for the JADE and Durham algorithms
respectively, and also with the results of
\cite{jets/moment_generating/Durham} for the Durham algorithm, there
calculated using moment generating functions as in section
(\ref{moment}).  The results presented here include, in addition to
these previous results, the non-abelian ($C_A$) terms for the JADE
algorithm, the five and six jet terms for the Durham algorithm, and
the abelian seven jet terms for both algorithms.  The abelian results,
the integrands for which are generated automatically, are limited by
the memory available to the computer.  The integrands for the
non-abelian results are put in by hand, and the results are therefore
limited by the number of relevant diagrams: the seven jet calculation
involves 32 such diagrams, compared with only 13 for six jets.

For the specific case of SU(3), $C_F=4/3$ and $C_A=3$.  The results of
equation (\ref{main_results}) then become
\begin{eqnarray}
R^J_2\openstrLsq 1.00 			\cr
R^J_3\openstrLsq 2.67 \strLsq		\cr
R^J_4\openstrLsq 3.33 \strLsq ^2	\cr
R^J_5\openstrLsq 3.45 \strLsq ^3	\cr
R^J_6\openstrLsq 3.24 \strLsq ^4	\cr
R^D_2\openstrLsq 1.00 			\cr
R^D_3\openstrLsq 1.33 \strLsq		\cr
R^D_4\openstrLsq 1.22 \strLsq ^2	\cr
R^D_5\openstrLsq 0.97 \strLsq ^3	\cr
R^D_6\openstrLsq 0.72 \strLsq ^4.
\end{eqnarray}
Again there is no clear pattern in the JADE results.  Although the
numerical coefficients do appear to be under control, it is difficult
to see how they might be resummed.  The root of the problem lies in
the necessity to compare all gluon pairings in the definition of the
phase space.  Here the Durham algorithm is an improvement, in that
fewer pairs of gluons need to be compared.

One further new result is the sequence of coefficients for a cascade
generated by a single gluon in the Durham algorithm:
\begin{eqnarray}
R^{Dg}_1\openfraction	1					\cr
R^{Dg}_2\openfraction	{1   \over 2}		\CAstrLsq	\cr
R^{Dg}_3\openfraction	{1   \over 6}		\CAstrLsq ^2	\cr
R^{Dg}_4\openfraction	{17  \over 360}		\CAstrLsq ^3	\cr
R^{Dg}_5\openfraction	{31  \over 2520}	\CAstrLsq ^4	\cr
R^{Dg}_6\openfraction	{691 \over 226800}	\CAstrLsq ^5.
\end{eqnarray}
The coefficients decrease less rapidly than the exponential series
owing to the increasing number of triple gluon vertices.

\section{Conclusion}

The jet fractions for the JADE and Durham algorithms have been
calculated up to six jets in the leading order in $\strong$ and
leading log order approximation.  The coefficients for the JADE
algorithm exhibit no easily discernable pattern, suggesting that they
may be difficult to resum.

I am most grateful to G.~P.~Salam and B.~R.~Webber for many useful
discussions.  I would also like to thank the CERN Theory Division for
hospitality while part of this work was carried out.

\clearpage
\appendix

\section{Program Details}
\label{program_details}

\subsection{Step Functions and Recursive Branching}

Where it appears as part of an integrand, a step function is defined
mathematically as a change in the limits of the integration, ie.
\begin{equation}
\int_{-\infty}^\infty\Theta(x-a)f(x)dx=\int_a^\infty f(x)dx.
\end{equation}
A more complicated situation arises when there are two step functions,
either of which may give the lower limit:
\begin{equation}
\int_{-\infty}^\infty\Theta(x-a)\Theta(x-b)f(x)dx=\cases{\int_a^\infty f(x)dx
& $a\ge b$,\cr \int_b^\infty f(x)dx & $b\ge a$.}
\end{equation}
The above result can be written in a combined form as
\begin{equation}
\int_{-\infty}^\infty\Theta(x-a)\Theta(x-b)f(x)dx=\Theta(a-b)\int_a^\infty
f(x)dx+\Theta(b-a)\int_b^\infty f(x)dx.
\end{equation}
In an integration over more than one dimension the constants $a$ and
$b$ can become functions of the remaining variables of integration.
The information already assumed about $a$ and $b$ is written in terms
of step functions and the integration over the next dimension is
therefore of the same form.  An integration in $n$ dimensions has been
branched recursively to two $n-1$ dimensional integrations.

The scheme works analogously for upper limits:
\begin{equation}
\int_{-\infty}^\infty\Theta(a-x)\Theta(b-x)f(x)dx=\cases{\int^a_{-\infty}
f(x)dx & $a\le b$,\cr \int^b_{-\infty} f(x)dx & $b\le a$.}
\end{equation}
If both limits are given by step functions, the step function giving
the lower limit must switch on before the step function giving the
upper limit switches off:
\begin{equation}
\int_{-\infty}^\infty\Theta(x-a)\Theta(b-x)f(x)dx=\Theta(b-a)\int_a^bf(x)dx.
\end{equation}
The most general case involves an unspecified number of both upper and
lower limits.  The answer contains one branch for every possible
pairing of an individual upper limit and an individual lower limit.
For each branch there are three sources of step functions in addition
to any step functions contained implicitly in $f(x)$:
\begin{itemize}
\item the lower limit must be greater than other possible lower limits,
\item the upper limit must be smaller than other possible upper limits,
\item the upper limit must be greater than the lower limit.
\end{itemize}

\subsection{Linked Lists}

At each stage of the integration it is not practical to predict the
number of terms that the polynomial component of the integrand will
contain, nor the number of step functions needed.  Therefore it is
essential to have a data structure that allows for an arbitrary number
of terms at each stage, for example a linked list.  The individual
terms are stored separately, each one with pointers to the next and
previous terms.  C \cite{C/book1,C/book2} supports dynamic memory
allocation: functions are included which can insert or remove terms at
any desired location, memory being taken from or returned to the pool
of free memory as required.  The recursive design of the integration
ensures that only one integrand need be stored for each dimension at a
time, thus minimising the memory use.

\subsection{Fractions and Euclid's Algorithm}

The initial integrand for each calculation consists numerically purely
of integers; real numbers do not occur.  This is the case for both the
JADE and Durham algorithms.  The integrand can therefore be expressed
purely using rational numbers at all stages of the calculation,
including the final answer.  The handling of fractions is done by a
separate program module; fractions are cancelled using Euclid's
algorithm for the highest common factor (HCF) of two integers
\cite{Euclid's_algorithm/reference,Euclid's_algorithm/explanation}, as
described below:

Let the two numbers whose HCF is to be found be $x_1$ and $x_2$, such
that $x_1>x_2$.  The procedure is best described schematically:
\begin{eqnarray}
x_1&=&m_1x_2+x_3 \cr
x_2&=&m_2x_3+x_4 \cr
&\vdots& \cr
x_n&=&m_nx_{n+1}.
\end{eqnarray}
At each stage the first number is divided by the second number, the
remainder being given by the third number.  Any common factor of $x_1$
and $x_2$ must divide $x_1-m_1x_2=x_3$, and is hence a common factor
of $x_2$ and $x_3$.  Any common factor of $x_2$ and $x_3$ must also be
a factor of $x_1$, and hence the HCF of $x_1$ and $x_2$ is the same as
the HCF of $x_2$ and $x_3$.  This argument is repeated for all
equations in the above series to show that the original HCF required
is also the HCF of $x_n$ and $x_{n+1}$: this is clearly $x_{n+1}$
itself.

\end{document}